\documentclass[10pt]{article}
\usepackage{graphicx}
    \usepackage{color}

\usepackage{epsfig}
\usepackage{latexsym} 
\usepackage{amsmath}

\begin{document}

\begin{centering}
\title{ Shannon  Entropy   and Herfindahl-Hirschman Index  as Team's Performance  and Competitive Balance Indicators in  Cyclist  Multi-Stage Races}
 
\vskip0.5cm
\author{ Marcel Ausloos $^{a,b,}$\footnote{ ORCID number : 0000-0001-9973-0019 
}  \\
\\$^a$ School of Business, University of Leicester,\\ Brookfield, Leicester, LE2 1RQ, UK\\   e-mail: ma683@leicester.ac.uk\\
 \\ $^b$ Group of Researchers Applying Physics in Economy and Sociology \\(GRAPES),  Beauvallon, rue de la Belle Jardini\`ere, 483/0021\\ Sart Tilman, B-4031, Angleur, Li\`ege, Belgium \\  e-mail: marcel.ausloos@uliege.be
 }
\end{centering}
\maketitle
\vskip -.5truecm
\begin{abstract} It seems that one cannot find  many papers relating entropy to sport competitions. Thus, in this paper, I  use  (i) the Shannon intrinsic entropy  ($S$) as an indicator of "teams sporting  value" (or "competition performance")  and (ii) the  Herfindahl-Hirschman index (HHi)  index as a  "teams competitive balance" indicator, in the case of  (professional) cyclist  multi-stage races. The 2022 Tour de France and 2023 Tour of Oman are used for numerical illustrations and discussion. The numerical values are obtained from classical and and new  ranking indices which measure the teams "final time", on one hand, and "final place", on the other hand, based on the "best three" riders in each stage, but also the corresponding times and places  throughout the race, for these finishing riders. 
 The analysis data  demonstrates that the constraint, "only the finishing riders count", makes much sense for obtaining a more objective measure of "team  value" and team performance", at the end of a multi-stage race. A graphical analysis allows to distinguish various team levels, with in each a  Feller-Pareto distribution, thereby pointing to self-organized processes.
In so doing, one hopefully better relates objective scientific measures to  sport team competitions, and,  besides, even proposes some paths to elaborate on forecasting through standard probability concepts.

\end{abstract}

\bigskip
   
 \section{Introduction}\label{Introduction}
 
Team ranking is a complex matter, as debatable as individual ranking, and it has been well known in modern times since Condorcet's paradox~\cite{Condorcet} and Arrow's impossibility theorem~\cite{Arrow1950}. This matter is highly relevant in sports competitions and is also a common practice in almost all aspects of social life, whether real or virtual. It is unnecessary to further point out the examples of university rankings, academic hiring, or promotion. Regardless of the method used, resulting ranks lead to prestige or contempt and are often accompanied by monetary rewards. Most of the time, in social life, ranking indicators encourage progress and help avoid mediocrity. 
  
Ranking is derived from a set of indicators, and there are many such indicators used in sports to rank athletes~\cite{Colllingwoodetal22} or teams~\cite{Stefani97,Wilson95}. In team competitions like football (soccer), basketball, hockey, rowing, etc., where the focus is sometimes on specific athletes, the quality of a team is determined based on various quantitative measurements of their ``quality'' or ``value.'' The ranking ``problem'' often arises due to the fact that competitions involve duals between athletes or teams, with incomplete round-robin formats. The case of cycling races is somewhat unique. It is widely accepted that cycling races are won by individual riders, but the role of the team is of crucial importance~\cite{Albert1991,Cabaud2022}. Quoting Cabaud et~al.~\cite{Cabaud2022}: 

\begin{quote}
  \emph{``A large proportion of cyclists in a race take part in support of another rider, meaning that they do not care about their personal result but instead try to help their team leader(s). Moreover, a team leader generally has one specific objective among a range of possible~ones.''}
\end{quote}
 
 In fact, cyclist races are quite different from other sport competitions, which emphasise the performance (physical or mental) of individual athletes, as in golf, boxing, escalade, tennis, judo,  climbing, chess, bridge, etc. In cyclist races, many teams, typically consisting of an equal number of athletes at the start of the race, compete in order to be rewarded through monetary prizes, somewhat regulated by UCI \footnote{UCI: Union Cycliste Internationale, e.g., $https://fr.uci.org/$}
  points, - or glory. The competition can be highly strategic: some athletes are geared toward a certain type of prowess, such as a time trial, sprint, mountain climbing, and other specialized cyclist races, not performed on roads. 

   For simplicity, races that take place ``off the roads'' are not considered in the discussion below, although it is evident that the concepts discussed here can be extended to such cases. Therefore, this paper specifically focuses on professional road cycling, particularly multi-stage races. 
 
 \subsection{Team Ranking}\label{TR}
 
 The literature about ranking teams, in sport, and in various  socio-economic matters, is huge, but  is much less abundant  about cycling teams. The papers of Sorensen,  in 2000,~\cite{sorensenrankingmethods} and Vaziri et~al., in 2018, ~\cite{Vaziri2018} appear to be the most pertinent ones about ranking teams  in (various) sports, because of their general viewpoints. 
 
 More pertinent to our concern, a 2020 paper by Ausloos~\cite{AusloosTdF2020}  studied data of professional cyclists from the Tour de France,  along the rank-size law method,  deducing financial considerations  from classical  UCI measures (time-based) team ranking.   Along a similar perspective, Ficcadenti et~al.~\cite{Ficcadenti2022}, in 2022, discussed the above-mentioned  rank-size law on a soccer competition in Italy, thereby observing  some regularities in team values within a ranking process. 
  
  One should notice that papers said to be  on ``performance ranking'' do not truly rank  teams (or athletes) from their score result at the finishing line, but rather refer to  body physiology and/or athletes' physical and biochemical conditions. These  aspects, outside the present discussion, are not considered here.
  
  For completeness, one should point to a non-scholarly website by  ProCyclingStats (PCS) as a data source where  interesting metrics are calculated; see    
$https://www.procyclingstats.com/info/point-scales$.

 \subsection{Competitive Balance}\label{CB}
 
 Fairness and equal chance opportunities are {\it a priori} pillars of  sport competition.  Hence, the literature on competitive balance is so huge,  stressing  qualitative and quantitative aspects, this paper mostly pertains to league activities rather than to teams point of view~\cite{Sanderson2002}. 
 
 The construction of rules to quantify competitive balance is of utmost importance, particularly for the lower-ranked teams whose main sources of revenue come from sponsorship and viewership~\cite{Rodriguezetal2015,RodriguezGutierrezFernandezBlanco2017}. This is why specific strategies are employed in cyclist races to ensure that teams, which may not have the most likely winners, have "attacking riders" right from the starting line. Therefore, a dynamic measure is required for cyclist races to adequately capture and reward competitive balance. However, such a measure does not appear to be currently available. Physiological performance is often linked to competitive balance as well~\cite{Cabaud2022}. For additional perspectives, readers may refer to economic-related papers, such as those focusing on the Tour de France~\cite{Bacik2021,AndreffMignot2022}.
 
  (N.B. During the writing of this paper, an interesting example of indirect imbalance occurred in the 2023 Giro when organizers provided helicopters to teams for transporting a few riders down from the Gran Sasso finish line after stage 7, but at a cost that only a few teams could afford~\cite{Lenten2015}). 

 \subsection{Shannon Entropy and Herfindahl–Hirschman Index}\label{SandHHi}
 
   The  cyclist teams sporting performance is based here on  the team's intrinsic entropy,  while the Herfindahl-Hirschman index  is used as the team's competitive balance indicator; both are mathematically defined here below,  and thereafter adapted to the studied sport cases. The literature abounds on both measures, but not much on cyclist races.
  
  \subsubsection{Shannon Entropy}\label{Slit}
  
  One can hardly comment in an original way on the use of entropy for measuring complex disorder~\cite{shannon}.  Alas, I have not found many papers relating entropy to sport {competition results, with the exception of~\cite{DaSilvaetal13,Silvaetal14,Silvaetal16}; however, their approach is somewhat different than that of the present framework}. This lack of papers in the literature is surprising since one could easily transform the occurrence of sport results  into probabilities, making it potentially useful for audacious forecasting and betting purposes, given appropriate scaling.  Thus, let it be considered that this paper is a pioneer contribution to the field. 
  
  {

In information science---particularly in scientometrics, in order to discuss uncertainty in measures, the concept of entropy ($S$) is a classical concept~\cite{shannon}, -- although it is sometimes misunderstood, or abusively misinterpreted. Its mathematical formulation reads as follows\footnote
{{In information science, entropy is usually defined through a log in base 2 in ``shannon units'';  in thermodynamics, the natural log is used, $ln\equiv log_2$: then, $S$ is given in ``nat units'', as it is here;  it is well known that a base change is easy, using the formula  
$log_a(b) = log_x(a) /log_x(b)$; $log_2(a) = ln(a) /ln(2);
 ln(2)=0.69315$.}} :
\begin{equation}\label{entropyeq}
  S = -\sum_{i=1}^N \frac{y_i}{\sum_j y_j} \cdot \ln \left(\frac{y_i}{\sum_j y_j}\right)
\end{equation}
where $\frac{y_i}{\sum_j y_j} \equiv p_i$ is the  probability of the number of occurrences,  $y_i$,  of the $i$-th event  among $N$ possible ones. 
 The  entropy maximum occurs when all  $p_i$ are equal to each other, i.e., when  there is no disorder :  $p_i=1/N$: 
 \begin{equation}\label{entropyMeq}
 S_{Max} = -  \ln  \left(\frac{1}{N}\right)
\end{equation}

Notice for the present study that the  entropy maximum occurs  when  each team wins as many stages as another; it may happen that some teams do not win  any stage. The team winning a stage might not be the team of the winning rider. It may also (often) happen that there are not enough stages such that each team wins one stage at least. 
The minimum entropy occurs when one team wins all $L$ stages.  
}
 
 \subsubsection{Herfindahl–Hirschman Index}\label{HHilit}
 
  In brief, measurements of  competitive balance are often based on ``Standard Deviation of Win Percentages''~\cite{Trandel2011}.  Other measures have been discussed~\cite{Humphreys2002,Triguero2019distance}. 
  
 
 Aside from the  ``Standard Deviation of Win Percentages''~\cite{Trandel2011}, the Herfindahl–Hirschman index is the most frequently used measure~\cite{OwenRyanW07,OwenOwen22}.  The $HHi$ is a concentration measure, which is typically used in business to emphasize monopolies by measuring the ``size'' of companies through their market share, hence providing some  numerical relationship between the firms and the competition they face. This measure  can be adjusted so that it reflects some aspect of competitive balance in sport by calculating the distribution of  points, won  through time or place, obtained by riders in a race competition.
  
Notice that the $HHi$ has been used on cyclist races such as  the Tour de France in
 ~\cite{AusloosTdF2020}, but with a focus on its ``market competition'' aspect, i.e., measuring teams' financial gains. 


Recall that the Herfindahl-Hirschman index ($HHi$), is  a measure of concentration~\cite{Hirschman}.  
It is usually applied   to describe company  sizes ({from} which the concentration  {is measured} with respect to the entire market):
a $HHi$  below 0.01 indicates a highly competitive index (in more usual language, from a portfolio point of view, a low $HHi$ implies a very diversified~portfolio).

When applied to the case of sport team ranking, the $HHi$ measure is proposed as an indicator of the level of fair competition among teams, rather than a measure of wealth concentration. Therefore, it serves as a competitive balance indicator. {(In political economy and finance, the $HHi$ in Equation~(\ref{HHIeq}) represents the sum of the squares of the market shares of the largest companies (traditionally, when expressed as fractions, with $N=50$). However, in the context of sports, typically $N<50$.)} Formally, the $HHi$ measure is defined as:
\begin{equation}\label{HHIeq}
HHi = \sum_{i}^{N} \left(\frac{y_i}{\sum_j y_j} \right)^2,
\end{equation}
where $y_i$ represents the number of wins by the $i$-th team, and $N$ is the appropriate number of teams. 

The higher the value of $HHi$, the smaller the number of teams with a large value of wins; in other words, $HHi$ is a measure of the number of the best-performing teams in a given competition. Thus, an increase in $HHi$ represents a decrease in competitive balance.

In conclusion of this  Introduction section, and in order to prepare the numerical illustrations and the subsequent discussion of findings and features, let it be mentioned that the data pertain to the Tour de France 2022 and the Tour of Oman 2023. In Section \ref{Data}, I explain where these illustrative data can be obtained, i.e.,  - from the official organizer websites.

\section{{Materials and Method}}

\subsection{Application to Multi-Stage Races}\label{MSRs}

Let us define the notations for clarity moving forward. Consider a race with $N$ teams and a total of $M$ riders participating in an $L$ stage race. Assuming that all teams start with an equal number of riders, there are $K = M/N$ riders per team initially. However, due to rider abandonments during certain stages ($s$), or riders not starting certain stages for reasons not relevant here, in a given team (\#), there are $K_{\#}$ riders who finish stage $s$, such that $M(s) = \sum_{\#} K_{\#}(s)$; 
$s = 1, \dots, L$.
For future reference, it is important to note that each rider ($i$) is assigned a bib number ($d_i$) by the race organizers at the start of the race. 

 On  the 2022 Tour de France (TdF), there were $L=$ 21 stages, and  $N=$ 22 teams.  Each team started with eight riders; therefore,  $M$ = 176.
 
 In the case of the 2023 Tour of Oman (ToO), there were  $N=$ 18 teams competing on $L=$ 5 stages.  Each team, with the exception of two of them,  started with seven riders;  therefore,  $M$ = 124.

\subsection{Data}\label{Data}
The relevant data can be obtained from the official websites of the race organizers or from media sources. {To conduct the necessary treatments and analyses, it is essential to download the results for the twenty-one stages and five stages, respectively. The official reports provided by the organizers come in various formats. In order to ensure consistency and facilitate subsequent analysis, certain data treatments have been performed manually.}

For instance, one can begin by accessing the stage results of the 2022 Tour de France from the official website $https://www.letour.fr/en/rankings$.
 The stages can be reviewed in reverse order starting from stage 21 and progressing backwards. Similarly, for the 2023 Tour of Oman, the official website $https://www.tour-of-oman.com/en$ 
can be visited to retrieve the final ranking in a single step through $https://www.tour-of-oman.com/en/rankings$
 Then, the previous stages can be accessed and reviewed in reverse order. 

\subsection{{Notations }}
The classical UCI counting goes as follows:  the time ($t_s^{(\#)}$) of the fastest 3 riders of a team $(\#)$ after a stage ($s$)  is aggregated, in order to give the ``team time'' for this given stage, say $T_s^{(\#)}$.  The ``final team time'' results from the aggregation of each ``team stage time'',  $  T_L^{(\#)}  = \Sigma_{s=1}^{L} \;\; t_s^{(\#)}$, - even though several of the riders, considered for the aggregated  ``team stage time'' might not have finished the whole race. 
 
 One may  also rank teams according to the  finishing place of the first three riders of a team at the end of each stage. (Notice that this measure differs from the so-called ``green jersey race'' for riders in the Tour de France).
 
 Let us introduce the relevant notations.
 Let such riders be at place  $ p^{(\#)}_{i,s}\;$,  with $i$~=~1, 2,  3. Thereafter, one can define another objective team ranking place after stage $s$  from $p_s^{(\#)}$, i.e., 
 \begin{equation}\label{psteameq}
p_s^{(\#)}  = \Sigma_{i=1}^{3}  \;\;  p^{(\#)}_{i,s}\;,
\end{equation}
and  calculate some 
\begin{equation}\label{PLteameq}
  P_L^{(\#)}  = \Sigma_{s=1}^{L}  \;\; p_s^{(\#)}  \;.
\end{equation}
at the end of the multi-stage race, for the final ranking, - according to the team placing at different stages, again, irrespectively of the involved riders.

 Notice that these $t$ and $p$ lists do not necessarily give the riders in the same order, due to the last (3) kilometer(s) neutralisation rule, allowing riders to have ``technical problems'', tire punctures, even falls, or to willingly stop racing,   along such a distance.

 Moreover, one can define the ``adjusted team final time'', $ A_L^{(\#)} $, calculated from  
 \begin{equation}\label{Ateameq}
  A_L^{(\#)}  = \Sigma_{j=1}^{3} \; \; t_{j,L}^{(\#)}
\end{equation}  
where, in Equation~(\ref{Ateameq}),  $j$ = 1, 2, 3 refers to the ``three  fastest''  riders of the team ${(\#)}$ \underline{having  completed all}  $L$ stages.  Thus,  $A_L^{(\#)}$  can only be so obtained at the end of the multistage race. 
 Let it be strongly emphasized  again that these three ``$j$'' riders might be quite different from the various three ``$i$'' riders having contributed to any  $t_s^{(\#)}$,  and hence to   $T_L^{(\#)}$.
 
 Similar to the above example, one can define   ``best team final place'' measures,  such as $P_L^{(\#)}$  and  $B_L^{(\#)}$,  based on the final place of the three ``best  riders'',   at the end of the race: 
$P_L^{(\#)}$ 
has been defined in Equation~(\ref{PLteameq}); recall that this  $P_L^{(\#)}$ measure refers to many different riders. To further refine the team ranking by considering only the riders ($j$) who successfully finish the race, ensuring a more comprehensive measure of team performance throughout the entire race, another quantity is defined as:

\begin{equation}\label{Bteameq}
B_L^{(\#)} = \sum_{j=1}^{3} p_{j,L}^{(\#)}
\end{equation}

This corresponds to $A_L^{(\#)}$ as defined in Equation~(\ref{Ateameq}). It is important to reiterate that in Equation~(\ref{Bteameq}), the index $j$ ranges from 1 to 3, representing the three best-placed riders of team $(\#)$ in the various stages, who have successfully completed all $L$ stages. 

 This leads to four different team rankings: in each case,  the best teams are those which have the lowest values of the four measured ``variables'', according to the ranking from measures in ascending order.
 
 One can easily plot and observe the distributions of such values. Moreover, their ranges being found to be finite, one can scale each result,  with respect to the respective sums. This allows for the definitions of {\it a posteriori} ``value probabilities'', which can be interpreted as the ``percentages of concentrations''.

These probabilities, denoted as $p_i$ to avoid confusion with $y_i$ in Equations~(\ref{entropyeq}) and (\ref{HHIeq}), can be incorporated into the definitions of $S$ and $HHi$ to obtain:

\begin{equation}\label{entropyeq2}
S = -\sum_{i=1}^N p_i \cdot \ln(p_i)
\end{equation}
which represents the classical formulation of Shannon entropy, and

\begin{equation}\label{HHIeq2}
HHi = \sum_{i}^{N} (p_i)^2
\end{equation}
which is also the standard way of measuring competitive balance through the "Standard Deviation of Win Percentages" ~\cite{Trandel2011}.
 
 \section{Results, Data Analysis}\label{Analysis} 
 
    \subsection{{Team Ordering Results}}

 {
  \subsubsection{2022 Tour de France (TdF)} \label{TdF}
 
 {Consider}   the 2022 Tour de France (TdF) results. There were $L=$ 21 stages, and 22~teams {(The team acronyms are those of the UCI notations)}. The final team ranking $T_L^{(\#)}$ leads to the hierarchy IGD, GFC, TJV, BOH, MOV,  $\dots$ TEN, ADC, LTS, QST.
In contrast, if the final ranking is based on the aggregated finishing time of the best three fastest riders that have finished the whole race, 
one obtains the  $A_L^{(\#)}$  ranking:  IGD, GFC, TJV, BOH, DSM, $\dots$ TEN, ADC, LTS, QST.

 Next, consider the  arriving place (instead of the time) of  the ``best placed'' three riders of a team in each stage.   The $P_L^{(\#)}$  values  lead to the following ranking: TJV, IGD, BOH, GFC, ARK,   $\dots$ TEN, BEX, ADC, LTS, QST, while
the  $B_L^{(\#)}$ hierarchy is IGD, TJV, BOH, GFC, AST, $\dots$ BEX, ADC,  TEN,  QST, LTS.

\subsubsection{2023 Tour of Oman (ToO)} \label{ToO}

 Consider this 2023 Tour of Oman (ToO) case. There were 18 teams competing in $L=$ 5 stages.
 The final team ranking $T_L^{(\#)}$,   is given by
  BOH, SOQ, ICW, ARK, $\dots$ , AST, HPM, TSG, ONT, while it is 
  BOH, ICW, SOQ, ARK, $\dots$ , AST, HPM, TSG, ONT,  for   $A_L^{(\#)}$ .

 The relevant $P_L^{(\#)}$   lead to BOH, SOQ, ARK, ACT, $\dots$ , JCL, HPM, TSG, ONT,
 but to  BOH, ICW, COF, UAD, $\dots$ , JCL, HPM, TSG, ONT  for $B_L^{(\#)}$.
 
}

 {
  \subsection{Statistical Characteristics} \label{Stat}
}
 
 A  summary of  the (rounded) main statistical characteristics  of the time and place indicator distributions, with the notations defined here above,  for  the 2022  Tour de France  and the 2023  Tour of Oman, for their respective  number of competing teams is given  in Table \ref{Table1statTAPLBTdFToO}. {It is observed from the Table that both investigated competitions are  on a quite different level from a sportive point of view: e.g., the number of teams and the number of stages are quite different.  From a purely statistical perspective, it is worth noting that the time or place measures, both with and without the constraint of considering only the finishing riders in the final team value or team performance measures, cover a wide range of scales. Further comments on this topic will be discussed in the next section Section \ref{Discussion}.

   \begin{table} 
   \caption{Summary of  (rounded) main statistical characteristics  of the time (in hours:minutes:seconds)  and place indicator distributions (see notations in the main text), for  the 2022  Tour de France  and the 2023  Tour of Oman;    $M$ is the number of competing teams, $L$ is the number of stages, in the respective cases.}
  
  \bigskip
  
 \setlength{\tabcolsep}{2.01mm}
\begin{tabular}[t]{cccccccccc}
   \hline\hline
  &\textbf{Min.}& \textbf{Max.} &   \textbf{Sum}& \textbf{Mean}    &   \textbf{St. Dev.}   &   \textbf{Skew.}    &   \textbf{Kurt.}     \\ 
   \hline\hline
& \multicolumn{7}{c}{2022   Tour  de   France; $M=22$;  {$L=21$}}  \\
 \hline
 $T_L^{(\#)}$	&	239:03:03	&	249:46:16	&	5361:45:14	&	243:42:58	&	3:01:12	&	0.43693	&	$-$0.70860	\\
$A_L^{(\#)}$	&	240:13:39	&	251:22:06	&	5401:26:40	&	245:31:13	&	3:09:10	&	0.08616	&	$-$0.86863	\\ 
$P_L^{(\#)}$	&	1242	&	4288	&	59753	&	2716.05	&	921.67	&	0.22809	&	$-$1.10963	\\ 
$B_L^{(\#)}$	&	1788	&	5899	&	86725	&	3942.05	&	1209.34	&	$-$0.01383	&	$-$1.12467	\\ 
\hline\hline
& \multicolumn{7}{c}{2023   Tour  of Oman; $M=18$; {$L=5$}}  \\ 
\hline
$T_L^{(\#)}$	&	59:55:12	&	61:43:42	&	1084:04:36	&	60:13:35	&	0:25:39	&	2.6427	&	6.66372	\\
$A_L^{(\#)}$	&	59:55:26	&	61:44:32	&	1085:04:31	&	60:16:55	&	0:26:50	&	2.2287	&	4.50161	\\
$P_L^{(\#)}$	&	267	&	1491	&	9131	&	507.28	&	296.60	&	2.2798	&	4.89609	\\
$B_L^{(\#)}$	&	357	&	1500	&	11245	&	624.72	&	285.83	&	1.9770	&	3.26794	\\ 
\hline
 \end{tabular} 
    \label{Table1statTAPLBTdFToO}
  \end{table}

Firstly, it is important to note that $A_L^{(\#)}$ consistently exceeds $T_L^{(\#)}$, even in races with few abandonments, such as the Tour of Oman. This observation suggests that the constraint of considering only the riders who complete the entire race is meaningful. Similarly, in terms of the "place value," $B_L^{(\#)}$ consistently surpasses $P_L^{(\#)}$. This clear trend emphasizes the significance of the constraint on the "finishing riders only" in obtaining a more objective measure of team value and team performance at the conclusion of a multi-stage race. 

 \section{
 Discussion 
}\label{Discussion}

The time and place distributions are displayed in Figures~\ref{FigPlot1timeTdFcubicforprint}--\ref{FigPlot4ToOPLBLcubicfitforprint}. The rank-time laws for the Tour de France, as shown in Figure~\ref{FigPlot1timeTdFcubicforprint}, exhibit a well-defined cubic form with a high coefficient of determination ($R^2 \ge 0.985$). It is worth noting that only the integer values on the x-axis hold significance in this context. On the other hand, for the Tour of Oman (ToO), which is a shorter race, the variations in team performance are relatively weak for most teams, except for the last three. As a result, fitting an empirical law is less meaningful. However, for the sake of completeness, it should be mentioned that a similar cubic fit with $R^2 \ge 0.985$ can be obtained if the weakest three teams (HPM, TSG, ONT) are excluded from the fitting process. 

Interestingly,  nevertheless, one can observe that aside from distinguishing two sub-distributions in the ToO results, observing again the data for TdF,     Figure~\ref{FigPlot3TdFPLBLnofitforprint},  one sees four possible sub-distributions, -  in fact, related to a team's ``UCI quality level''.

\begin{figure}
\includegraphics[width=0.68\textwidth]{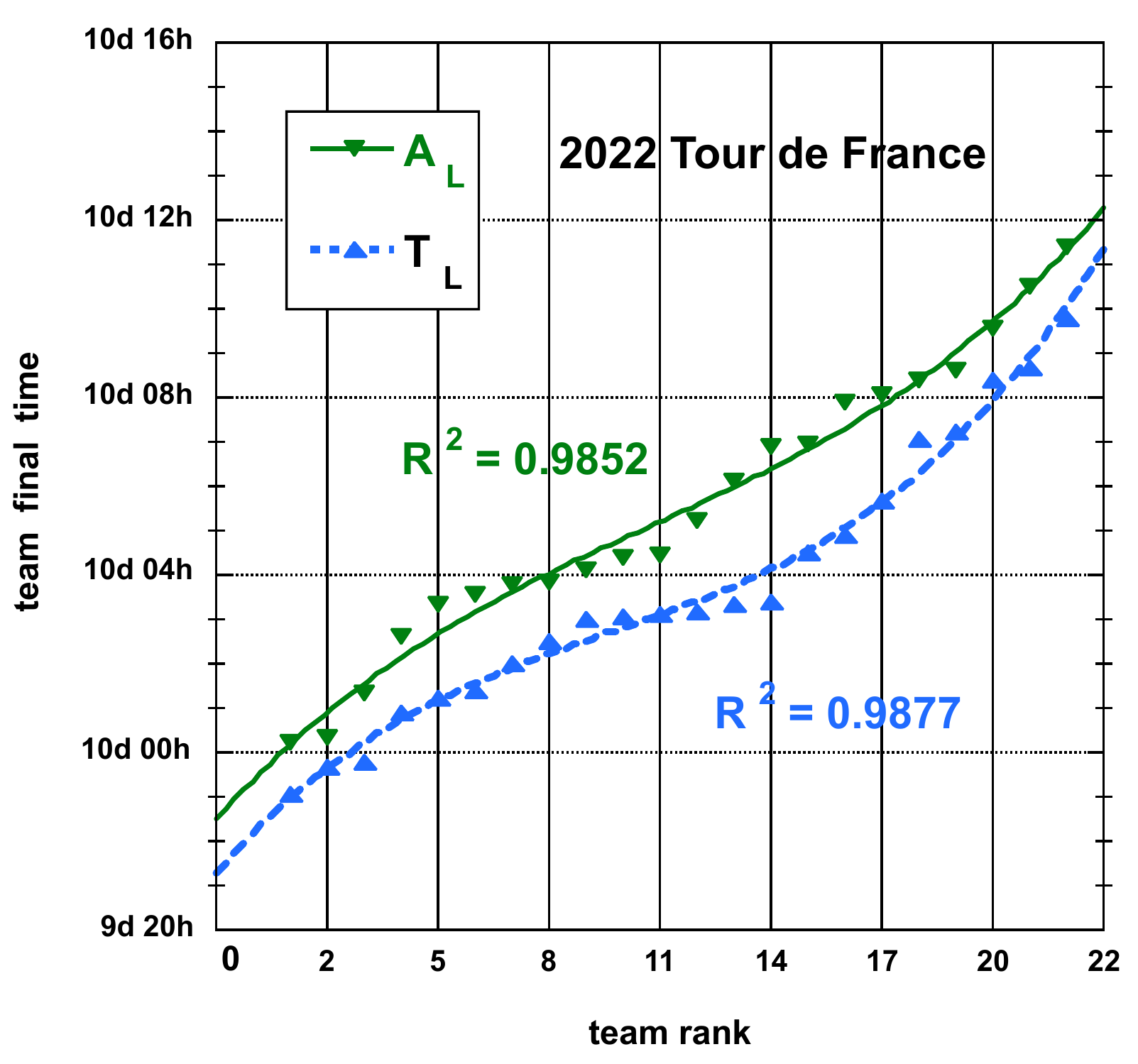}
\caption{
Best-fit curve to an empirical cubic function of the  distributions of final time $T_L$ and adjusted final time $A_L$, both defined in  the main text,  of the 22 teams having competed on the Tour de France 2022; the \emph{y}-axis scale is  in days:hours.
}\label{FigPlot1timeTdFcubicforprint}
\end{figure}


\begin{figure} 
\includegraphics[width=0.68\textwidth]{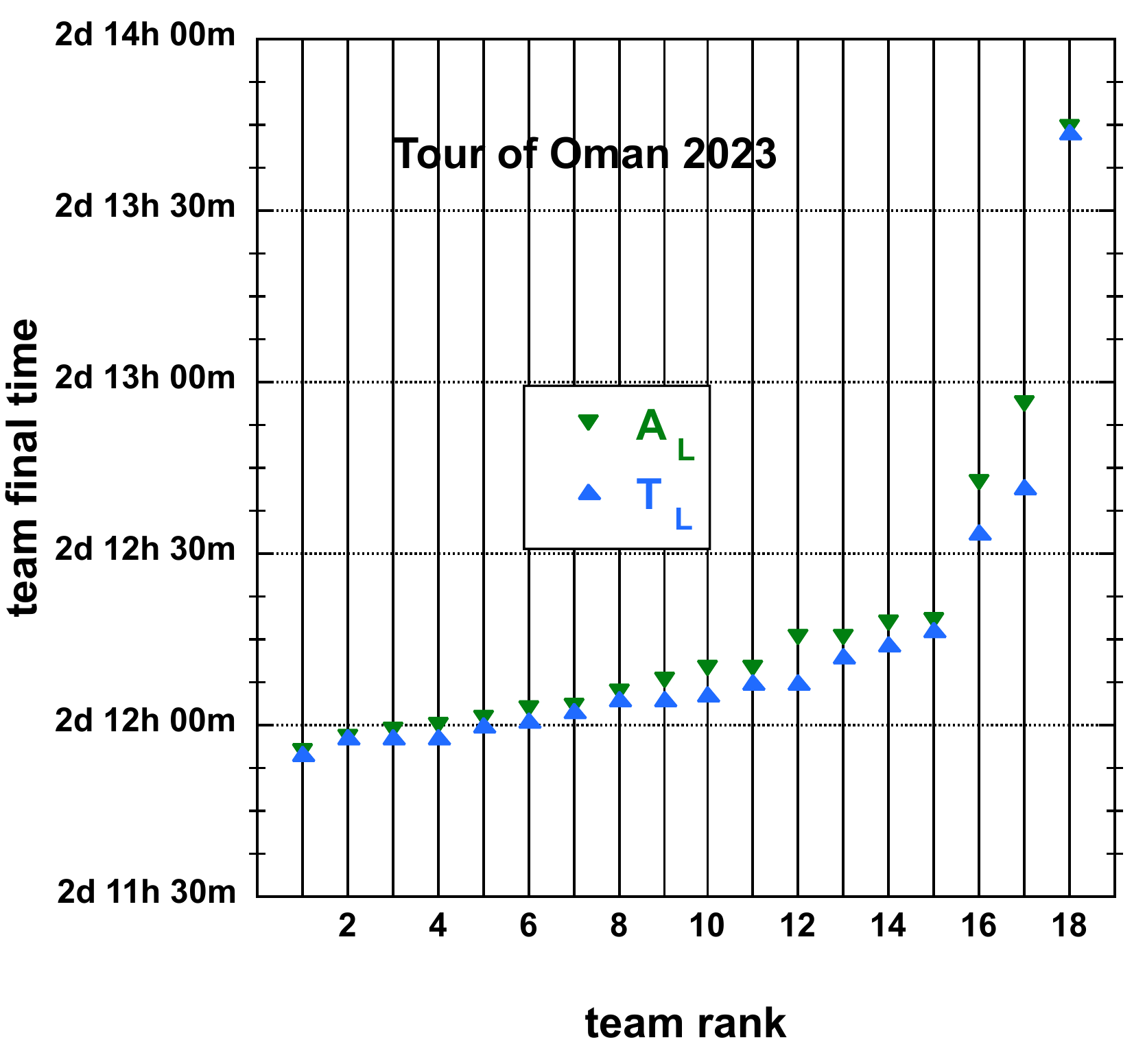}
\caption{
  Distributions of final time $T_L$ and adjusted final place $A_L$, both defined in  the main text,  of the 18 teams having competed on the Tour of Oman 2023; in order to emphasize the 2  sub-distributions, the best-fit curve to an empirical cubic function is  not shown; the \emph{y}-axis scale is  in days:hours:minutes.
}\label{FigPlot2ToOTLALnofitforprint}
\end{figure}

\begin{figure} 
\includegraphics[width=0.68\textwidth]{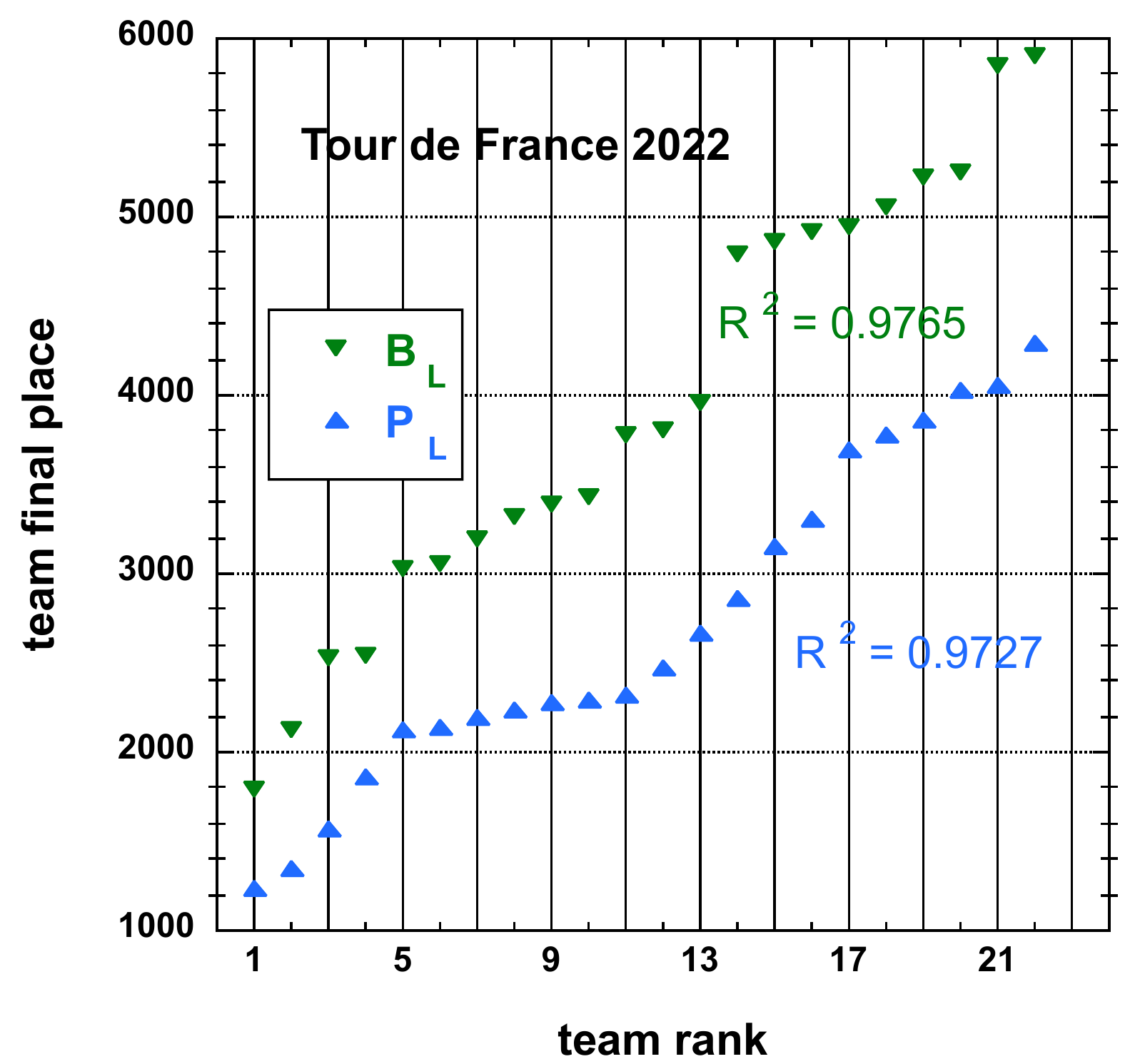}
\caption{
  Distributions of final place $P_L$ and adjusted final place $B_L$ measures, both defined in  the main text,  of the 22 teams having competed on the Tour de France 2022; in order to emphasize the four  sub-distributions, the best-fit curve to an empirical cubic function is  not shown, but the resulting $R^2$, if the whole fit was completed,  is given for  information.
}\label{FigPlot3TdFPLBLnofitforprint}
\end{figure}

 \begin{figure} 
\includegraphics[width=0.68\textwidth]{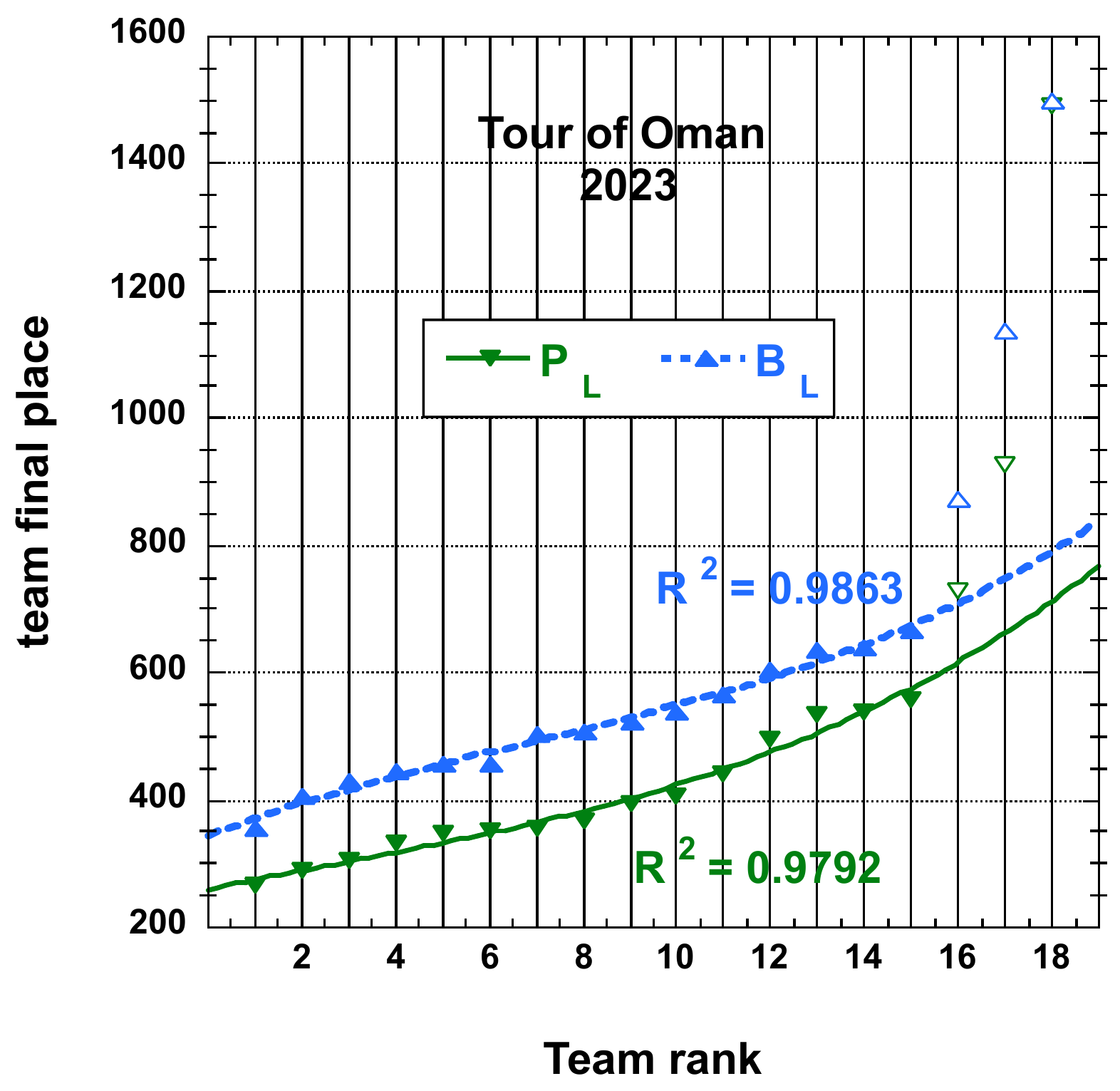}
\caption{
  Distributions of final place $P_L$ and adjusted final place $B_L$ measures, both defined in  the main text,  of the 18 teams having competed on the Tour of Oman 2023; the best-fit curve to an empirical cubic function is made on the first 15 teams only.
}\label{FigPlot4ToOPLBLcubicfitforprint}
\end{figure}

Concerning the entropy data, see Figures~\ref{FigPlot5TdFplnpTLALcubicforprint}--\ref{FigPlot8PLBLToOforprint}. Best fits to simple polynomials can be attempted. A best-fit curve to an empirical cubic function of the  team entropy  derived from the final time $T_L$ and adjusted final time $A_L$, both defined here above,  distributions of the 22 teams having competed on the Tour de France 2022 is displayed  on Figure~\ref{FigPlot5TdFplnpTLALcubicforprint}.  A similar fit for  the  team entropy, but derived from the final place $P_L$ and adjusted final place $A_L$ measures,  is shown on Figure~\ref{FigPlot6ToOTLALforprint}.

\begin{figure} 
\includegraphics[width=0.68\textwidth]{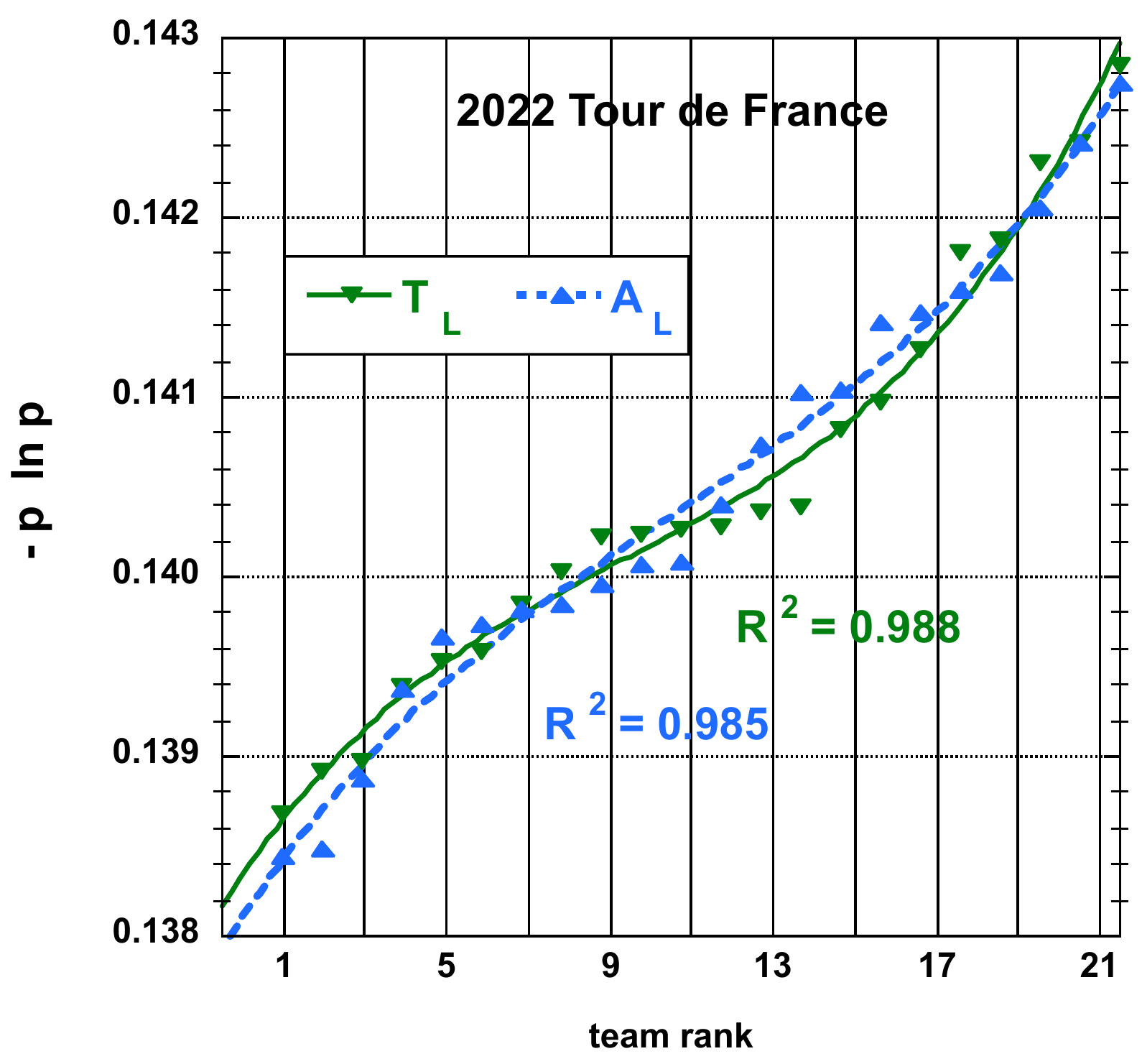}
\caption{
Best-fit curve to an empirical cubic function of the  team entropy  derived from the final time $T_L$ and adjusted final time $A_L$ measures, both defined in  the main text,  distributions of the 22 teams having competed on the Tour de France 2022.}\label{FigPlot5TdFplnpTLALcubicforprint}
\end{figure} 

\begin{figure} 
\includegraphics[width=0.68\textwidth]{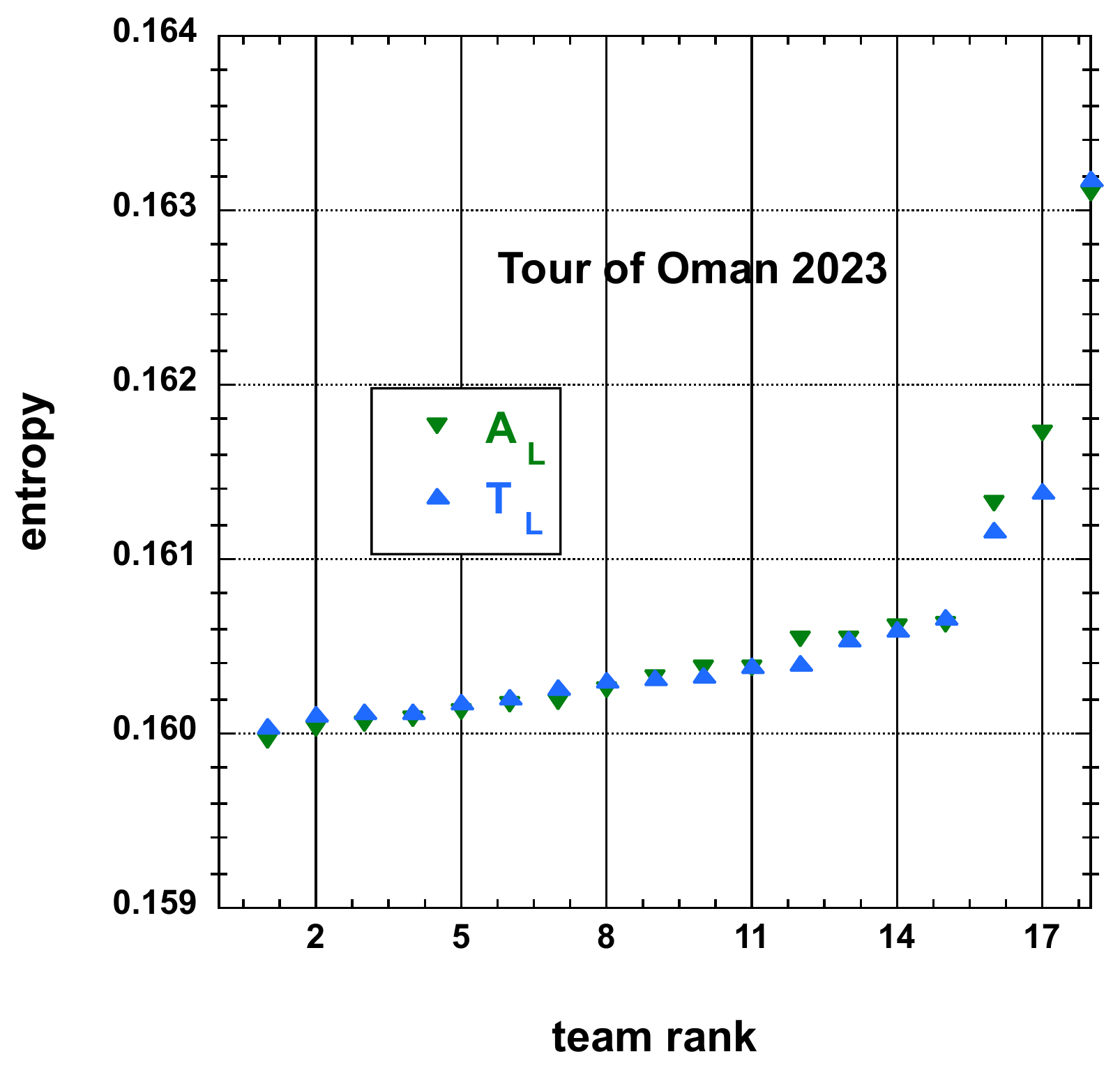}
\caption{
Team entropy  derived from the final time $T_L$ and adjusted final time $A_L$ measures, both defined in  the main text,  distributions of the 18 teams having competed on the Tour of Oman 2023.}\label{FigPlot6ToOTLALforprint}
\end{figure}

 The entropy  derived from $P_L$ and $B_L$  distributions for Tour de France 2022 and that
 for $P_L$ and $B_L$ for Tour of Oman 2023 are found on Figure~\ref{FigPlot7TdFPLBLforprint}  and on Figure~\ref{FigPlot8PLBLToOforprint}, respectively, with the best fit to a cubic empirical function.

\begin{figure} 
\includegraphics[width=0.68\textwidth]{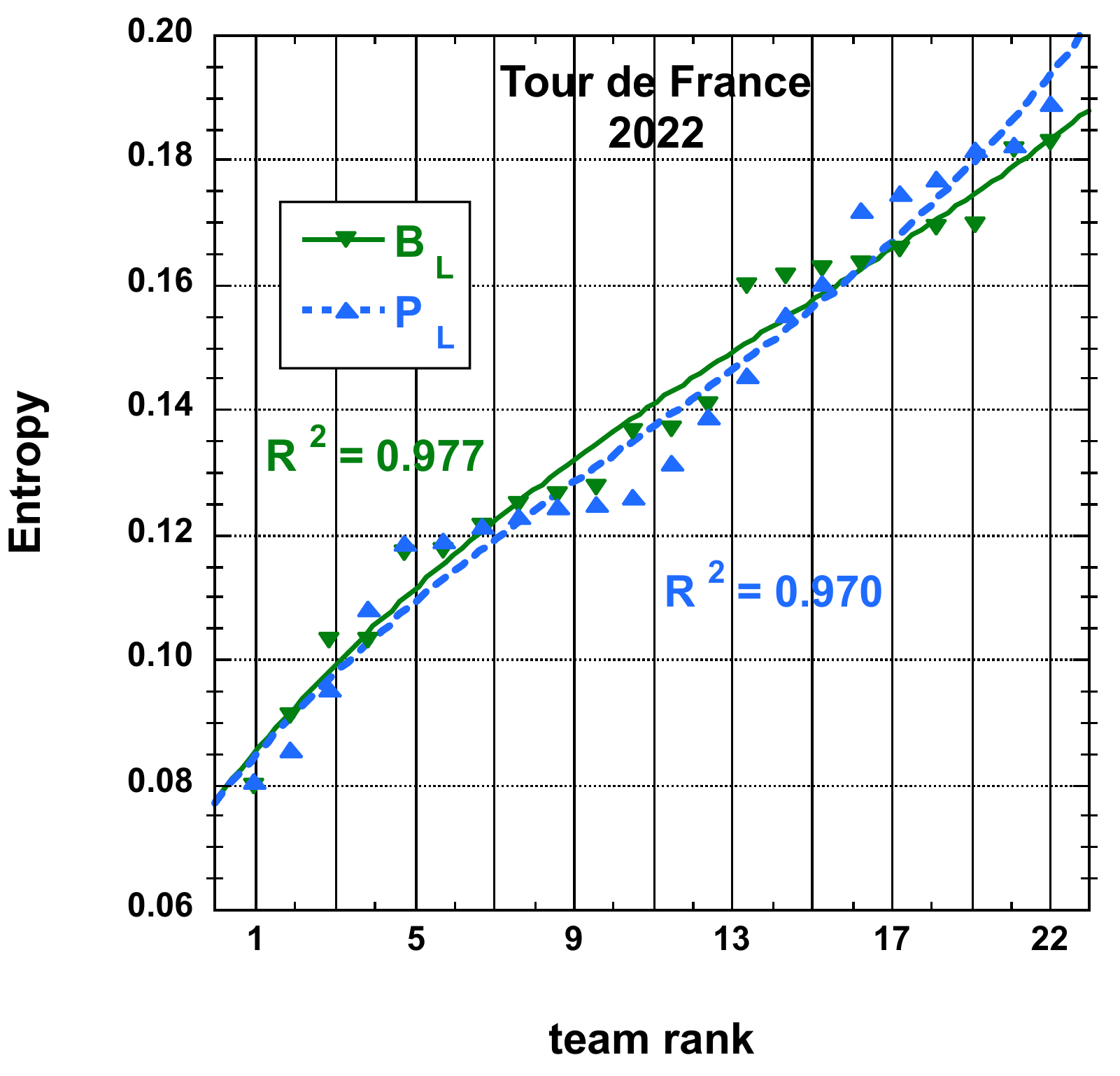}
\caption{
Best-fit curve to an empirical cubic function of the  team entropy  derived from the final place distributions,  $P_L$ and adjusted final place $B_L$ measures, as defined in  the main text,  for the 22 teams having competed on the Tour de France 2022.}\label{FigPlot7TdFPLBLforprint}
\end{figure}
\begin{figure} 
\includegraphics[width=0.68\textwidth]{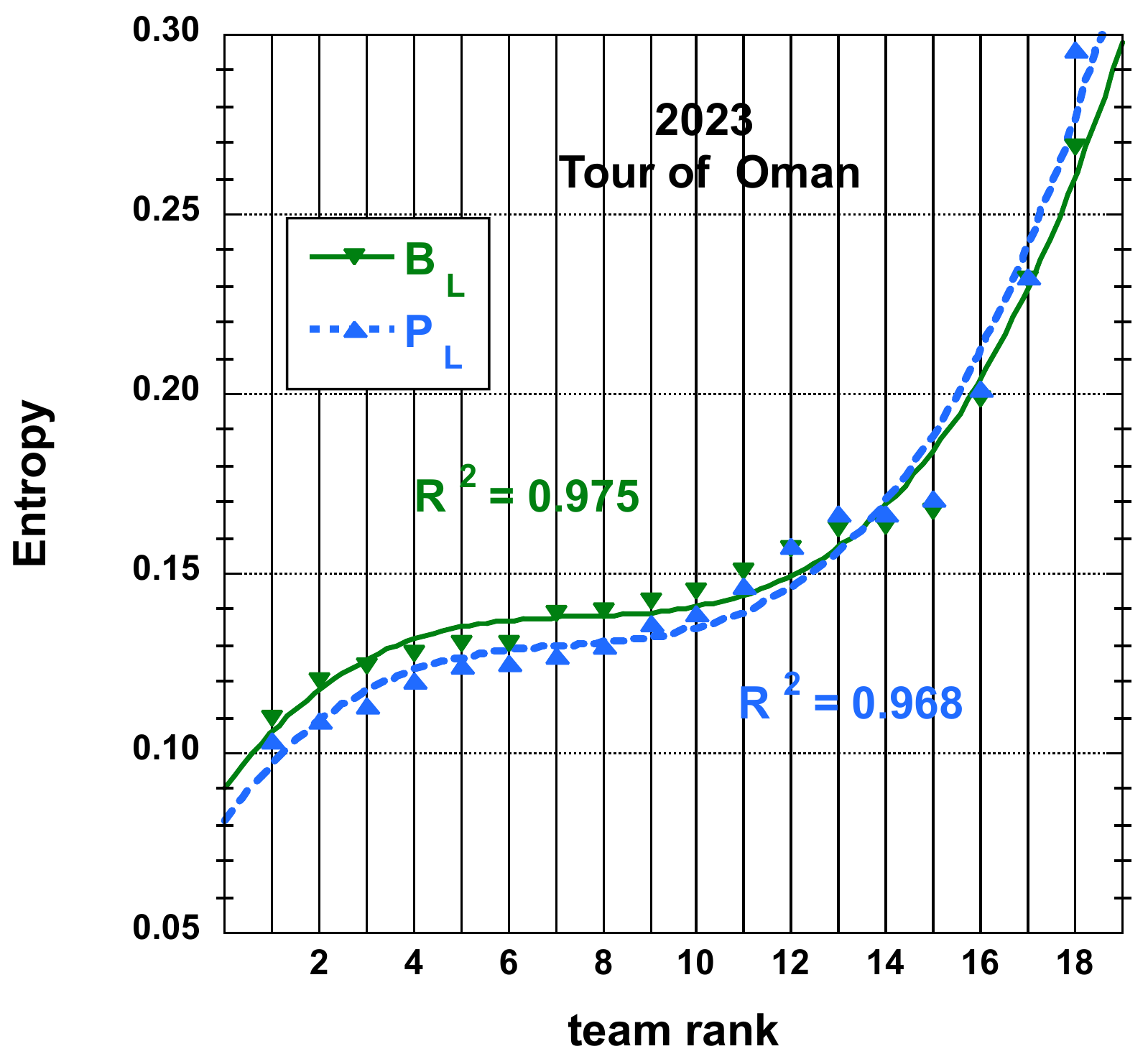}
\caption{
Best-fit curve to an empirical cubic function of the  team entropy  derived from the final place distributions,  $P_L$ and adjusted final place $B_L$ measures, as defined in  the main text,  for the 18 teams having competed on the Tour of Oman 2023.}\label{FigPlot8PLBLToOforprint}
\end{figure}

 Concerning the the  team $HHi$  values, they can also be displayed according to the team rank, both in  increasing order. The
 best-fit curve to an empirical cubic function for the  team $HHi$  derived from the final time $T_L$ and adjusted final time $A_L$, both as defined here above,  distributions of the 22 teams having competed on the Tour of France 2022 and for the 18~teams having competed on the Tour of Oman 2023  are found in Figure~\ref{FigPlot9p2HHiTdFTLALforprint} and in Figure~\ref{FigPlot10p2HHiToOTLALforprint}, respectively.  The corresponding $R^2$ is reported.

\begin{figure} 
\includegraphics[width=0.9\textwidth]{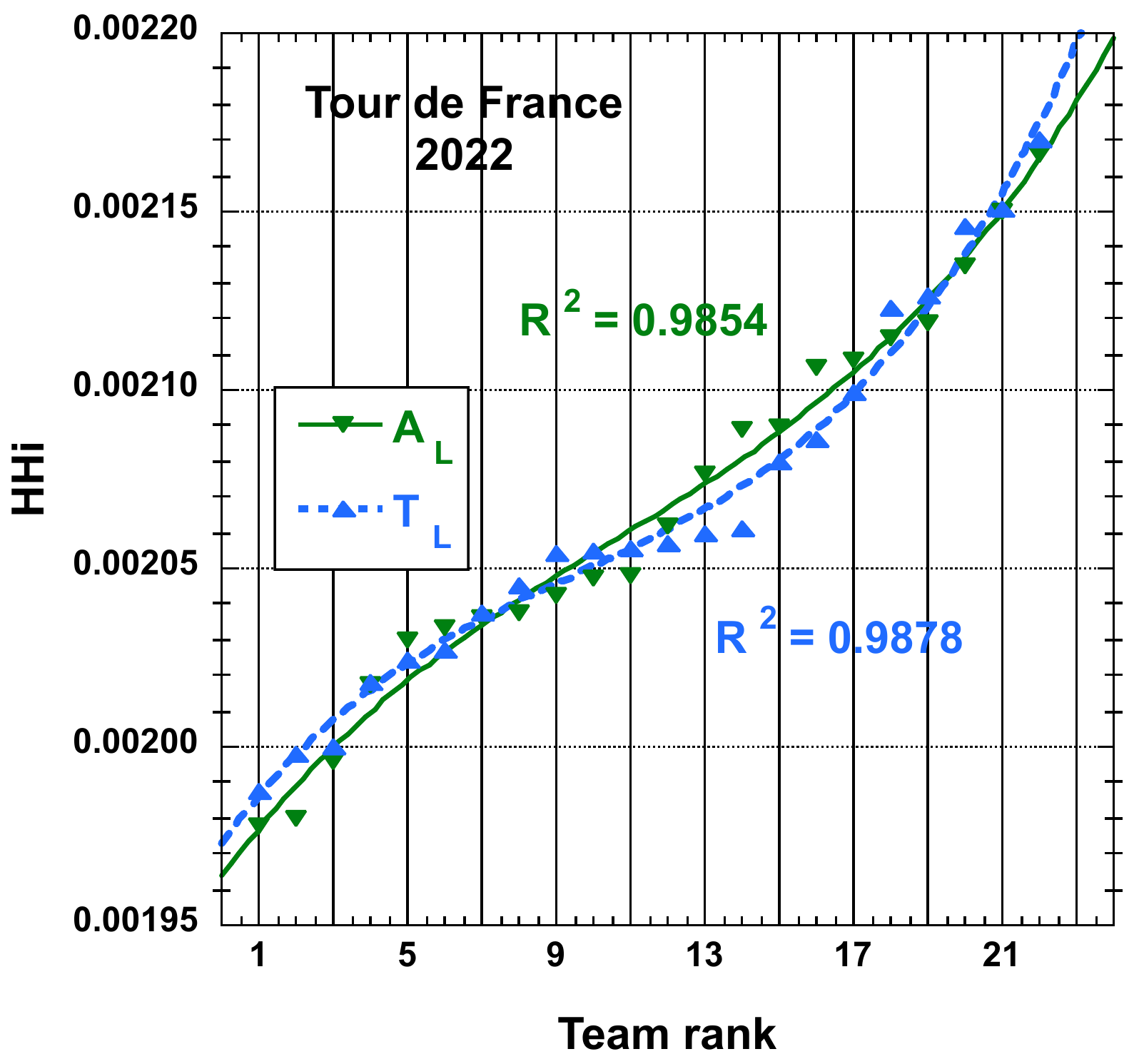}
\caption{
Best-fit curve to an empirical cubic function of the  team $HHi$  derived from the final time $T_L$ and adjusted final time $A_L$ measures, both defined in  the main text,  distributions of the 22 teams having competed on the Tour of France 2022.
}\label{FigPlot9p2HHiTdFTLALforprint}
\end{figure}

\begin{figure} 
\includegraphics[width=0.9\textwidth]{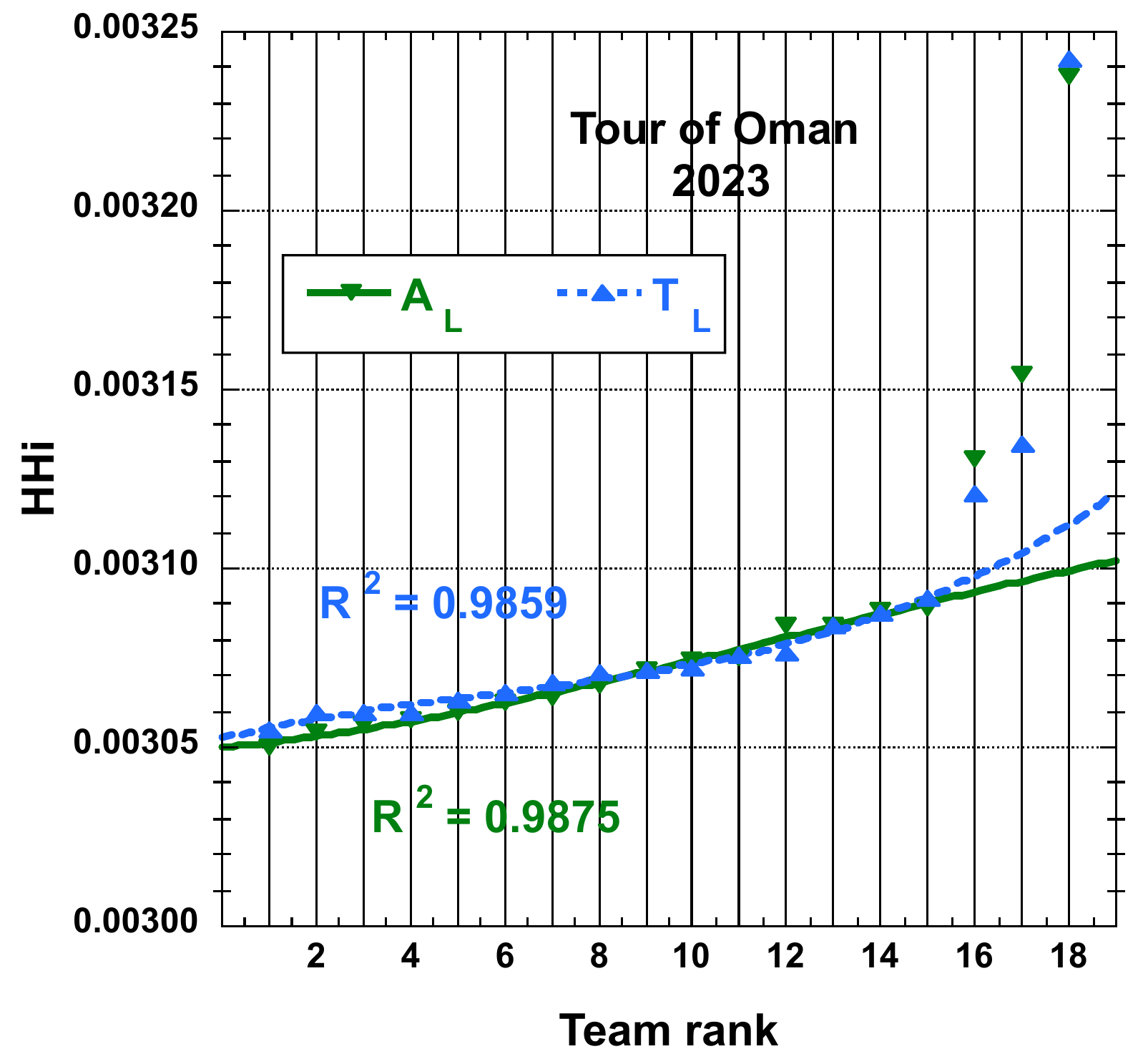}
\caption{
Best-fit curve to an empirical cubic function of the  team $HHi$  derived from the final time $T_L$ and adjusted final time $A_L$ measures, both defined in  the main text,  distributions of the 18 teams having competed on the Tour of Oman 2023; the fit is on the first 15 teams result.}\label{FigPlot10p2HHiToOTLALforprint}
\end{figure}

The corresponding displays for the $HHi$-rank distributions derived from the final place distributions,  $P_L$ and adjusted final place $B_L$, as defined here above,   for the 22 teams having competed on the Tour de France in 2022 and  for the 18 teams having competed on the Tour of Oman in 2023 are found 
 in Figure~\ref{FigPlot11p2HHiPLBLTdFforprint} and in Figure~\ref{FigPlot12HHiToOPLBLforprint}, respectively. The best-fit curve to an empirical cubic function  is given with the corresponding $R^2$.

\begin{figure} 
\includegraphics[width=0.9\textwidth]{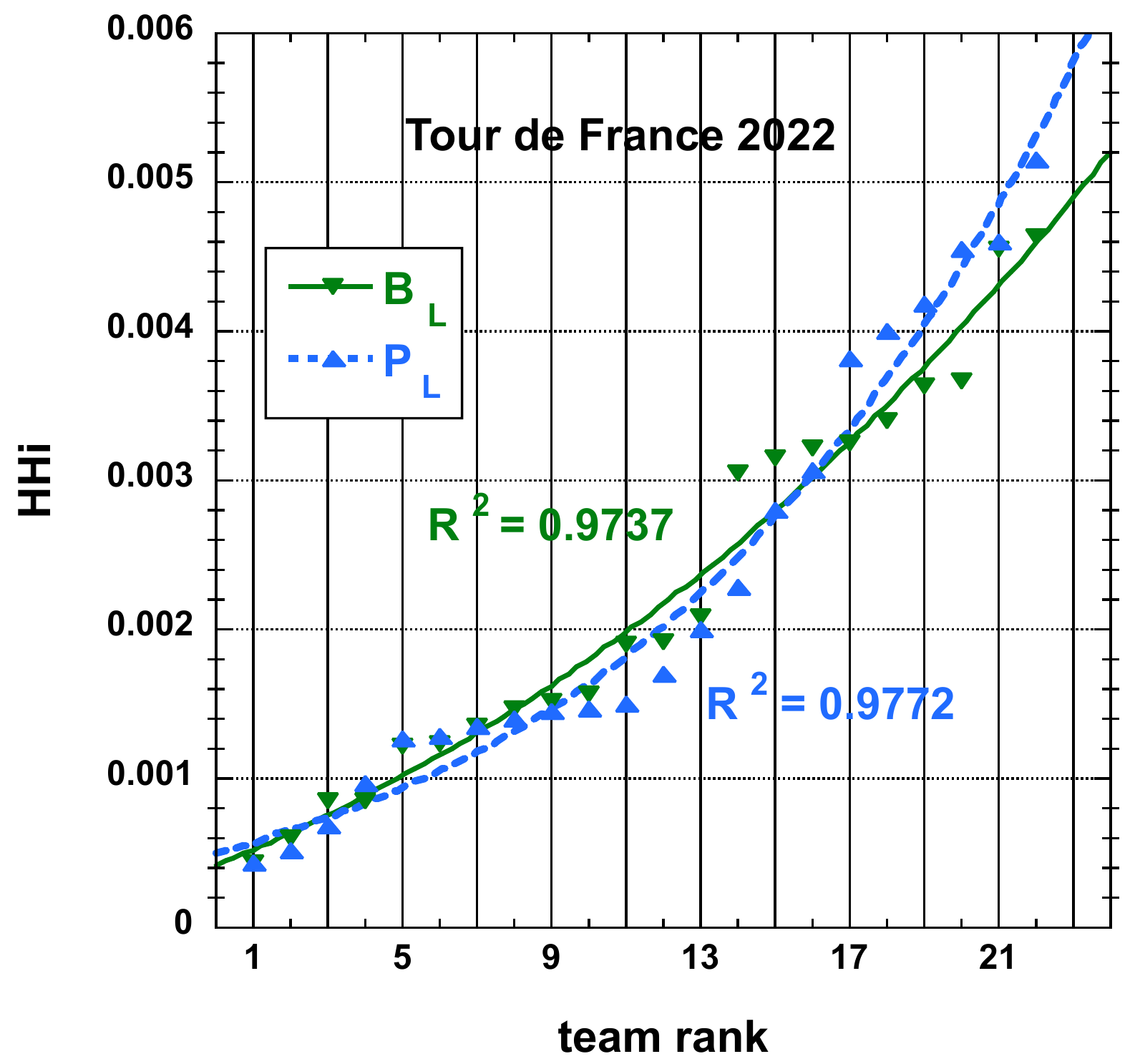}
\caption{
Best-fit curve to an empirical cubic function of the  team $HHi$ derived from the final place distributions,  $P_L$ and adjusted final place $B_L$ measures, as defined in  the main text,  for the 22 teams having competed on the Tour de France 2022.}\label{FigPlot11p2HHiPLBLTdFforprint}
\end{figure}

N.B. Let us still mention that the (cubic polynomial)  fits are reasonable on the first 15~teams results only, rather than on the whole 18 bunch for the Tour of Oman cases.
\begin{figure} 
\includegraphics[width=0.9\textwidth]{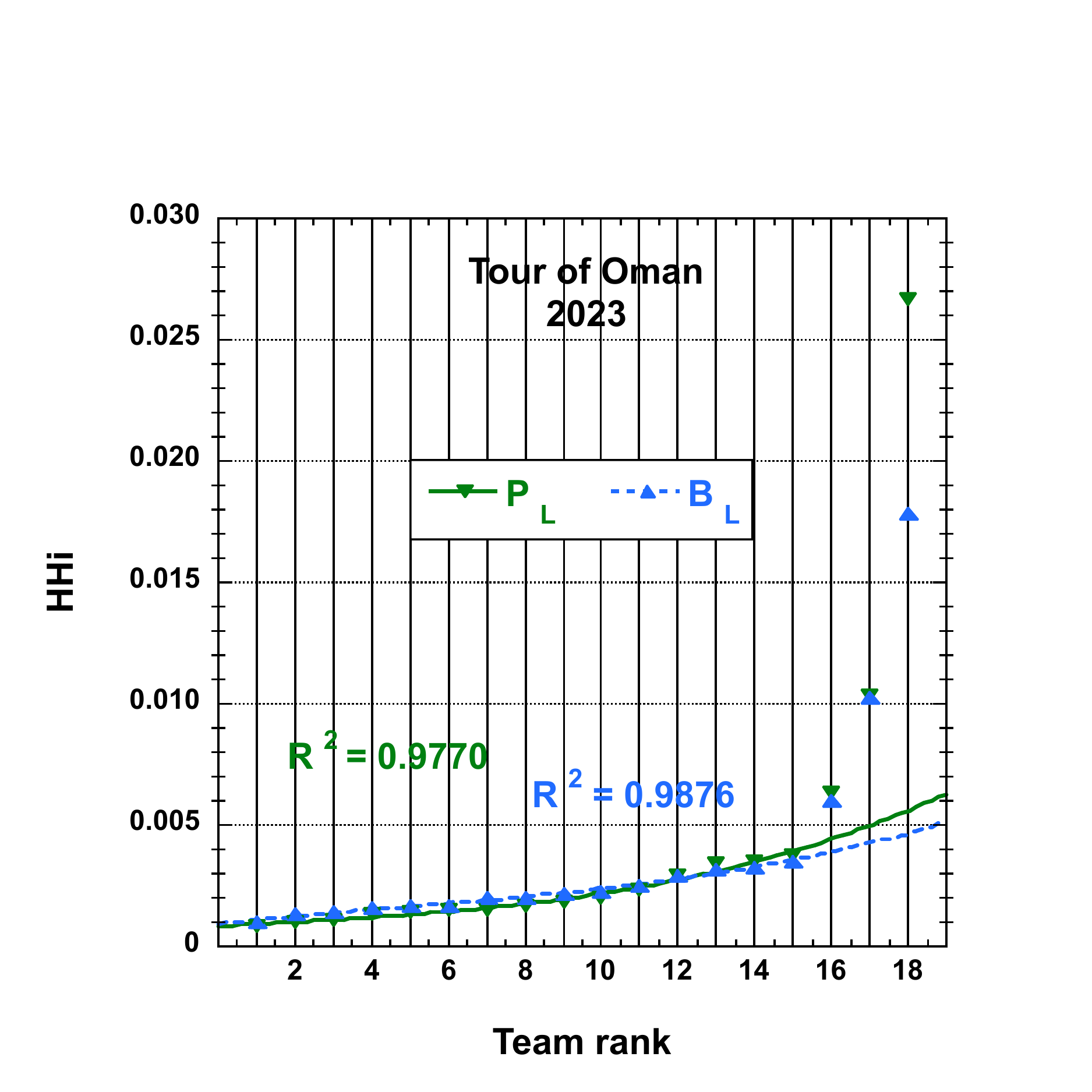}
\caption{
 Best-fit curve to an empirical cubic function of the  team $HHi$ derived from the final place distributions,  $P_L$ and adjusted final place $B_L$ measures, as defined in  the main text,  for the 18 teams having competed in the Tour of Oman 2023; the fit is on the first 15 teams' results.}\label{FigPlot12HHiToOPLBLforprint} 
\end{figure}

 These smooth variations are markedly different from classical rank-size laws analyses based on $y_r = a\; r^{-\alpha}$~\cite{MAJAQM}. However,  one could admit that other functions with a small number of parameters could be used in appropriately managing the $x$- and $y$-axes scales and range. Using the two-parameter   ($\kappa$, $\chi$) form
 \begin{equation} \label{Lavalette2} 
y(r)= \kappa\; \big[\frac{N\;r}{N-r+1} \big] ^{-\chi}  
\end{equation} would be an example of this. 

On a semi-log plot, Equation~(\ref {Lavalette2}) with $\chi\le0$, gives  a flat   N-shape ``$noid$'' function near its inflection point, allowing for various convex and concave data display shapes. Some rewriting~\cite{MAJAQM,ausloos2014twoexpoPRErelig}, defining $u= r/(N+1)$,  leads to
\begin{eqnarray}\label{Lavalette3}
\;\;  y(r)= \hat{\Lambda}\;   u^{-\phi}\;(1-u)^{+\psi}\;
\end{eqnarray}
which, in the case of $\phi >0$ and $\psi <0$,  is the Feller-Pareto function~\cite{tahmasebiAMS2010shannon}, which is associated with self-organized processes~\cite{eliazarcohen2011universalPhA,cerquetiGRMA2018investigating}. Exploring these features for different scenarios involving the number of stages, teams, and riders would be an interesting avenue for future investigation. It is clear that such races exhibit both endogenous and exogenous self-organization dynamics. 

Finally, a deduction can stem from observing different hierarchies through the various indicators: one can imagine various ways  to motivate teams through financial rewards, or media publicity,  based on  the above indicators.

 \section{Conclusions}\label{Conclusions}
 
In this paper, the author proposes the use of the intrinsic entropy ($S$) as an indicator of ``teams sporting value'' or ``performance'' and the Herfindahl-Hirschman index ($HHi$) as a measure of “teams competitive balance” in the context of professional cyclist multi-stage races.

The motivation for this study stems from the observation that there are relatively few papers linking the concept of entropy to sport competition results, with only a few notable exceptions. This is surprising considering that sport results can be easily transformed into probabilities, which have implications for forecasting and betting purposes. However, such analysis requires extensive computer work, including the downloading of appropriate results and scaling them according to the number of teams, number of stages, stage difficulties, stage lengths, and other relevant factors. 
 
 {Moreover,   even though many papers have been concerned with the concept of competitive balance in sports, recall the review in~\cite{OwenRyanW07}, the present study
   differs from  $HHi$-based theoretical, empirical, or simulation studies, as in  the very recent~\cite{OwenOwen22}. Most of these papers  take into account  different leagues' competitions, in different season lengths with different competitions rounds. Some relation can be tied to the present paper studying two different  lengths and difficulties of multi-stages races. However,  there are major differences with respect to other works on $HHi$ in sport stem in the type of competitions which are examined.  The papers in the literature mainly look at duel competitions. This is very different from professional cyclist races which involve many teams in a single event. 

As a second contribution aimed at objectively measuring team value and team performance, a logical constraint is introduced for the classical measures based on the time or place of riders in a team. It is mandated that these measures be solely based on the riders who finish the multi-stage race. This constraint leads to the introduction of new indicators, which are compared to the classical ones. The classical measures aggregate values from various riders in different stages but fail to capture the crucial contributions of the best race-finishing team members.
 
 In the present study, we utilize the 2022 Tour de France and the 2023 Tour of Oman as numerical illustrations and for discussion purposes. The former is a renowned long race featuring top professional teams, while the latter is a more modest race involving teams and riders who may be less well-known. Despite these differences, both races exhibit similar final characteristics. The numerical values are derived not only from new ranking indices that measure the teams' ``final time'' and ``final place'' based on the top three riders in each stage but also from the aggregated final time or final place of the team's best three finishing riders at the end of the race. The distributions of these values reveal distinct team levels through sub-distributions. Furthermore, these distributions are reasonably well-fitted by cubic (empirical) functions, reminiscent of the Feller-Pareto distribution, which suggests the presence of constrained self-organizational processes.
 
 A conclusive final point pertains to the empirical findings:  the intrinsic Shannon entropy  ($S$) appears to be a {\it bona fide}  indicator of ``teams sporting  value'' (or ``performance''), beyond its traditional role in measuring disorder. Furthermore,  the analysis reveals an unexpected yet understandable phenomenon of ``clustering of teams'' through data observation. On the other hand, the Herfindahl-Hirschman index ($HHi$) provides a clear indication of ``teams competitive balance'' processes. Overall, these empirical findings highlight the effectiveness of these indicators in capturing important aspects of team performance and competitive dynamics. 
 
\vspace{6pt} 
Funding: This research received no external funding.
\bigskip

Conflicts of interest: The author declares no conflict of interest.

\bigskip
Institutional review: ~ Not applicable.

\bigskip

Data availability: Not applicable.
\bigskip

{\bf Acknowledgments: } Comments by reviewers and editors have much  forged the structure and content of this paper.
\bigskip

 
 \end{document}